# The Dependency of the Cepheid Period-Luminosity Relation on Chemical Composition

M. Romaniello, F. Primas, M. Mottini, S. Pedicelli, B. Lemasle, G. Bono, P. François, M.A.T Groenewegen and C.D. Laney

*European Southern Observatory, Karl-Schwarzschild-Strasse 2, 85748 Garching bei München, Germany*
*Università of Roma Tor Vergata, Department of Physics, via della Ricerca Scientifica 1, 00133 Roma, Italy*
*Université de Picardie Jules Verne, Faculté des Sciences, 33 rue Saint-Leu, 80039 Amiens Cedex 1, France*
*INAF - Osservatorio Astronomico di Roma, via Frascati 33, 00040 Monte Porzio Catone, Italy*
*Observatoire de Paris-Meudon, GEPI, 61 avenue de l'Observatoire, 75014 Paris, France*
*Royal Observatory of Belgium, Ringlaan 3, 1180 Brussels, Belgium*
*West Mountain Observatory, Utah*

**Abstract.** The dependency of the Cepheid Period-Luminosity Relation on chemical composition at different wavelengths is assessed via direct detailed abundance analysis of Galactic and Magellanic Cepheids, as derived from high resolution, high signal-to-noise spectra. Our measurements span one order of magnitude in iron content and allow to rule out at the ~9 sigma level the universality of the Period-Luminosity Relation in the V band, with metal rich stars being fainter than metal poor ones by ~0.3 mag. The dependency is less pronounced in the K band. Its magnitude and statistical significance decisively depend on detailed distance measurements to individual stars, as inferred via the Infrared Surface Brightness Method.

## INTRODUCTION

The debate on the role played by the chemical composition on the pulsation properties of Cepheids is still open, with different theoretical models and observational results leading to markedly different conclusions. From the theoretical point of view pulsation models by different groups lead to substantially different results (e.g. [1], [2], [3]). On the observational side, the majority of the constraints come from indirect measurements of the metallicity, mostly in external galaxies, such as oxygen nebular abundances derived from spectra of H II regions at the same Galactocentric distance as the Cepheid fields (e.g. [4], [5], [6]).

The novelty of our approach consists in its simplicity: the direct and accurate measurement of metal abundance of a large sample of Cepheids spanning a factor of ten in metallicity and for which distances and BVJK photometry are available. Please refer to [7] for the full details of our analysis and our results.

## THE SAMPLE

We have observed a total of 68 Galactic and Magellanic Cepheid stars. The spectra of the 32 Galactic stars were collected at the ESO 1.5 m telescope on Cerro La Silla with the Fibrefed Extended Range Optical Spectrograph (FEROS, [8]). The spectra of the 22 Cepheids in the LMC and the 14 Cepheids in the SMC were obtained in service mode at the VLT-Kueyen telescope on Cerro Paranal with the UV-Visual Echelle Spectrograph (UVES, [9]).

The selected Cepheids span a wide period range, from 3 to 99 days. For the Galactic Cepheids we have adopted periods, optical and NIR photometry from [10], [11], [12], [13], [14]. For the Magellanic Cepheids we have adopted periods, optical and NIR photometry from [10].

The 2D raw spectra were run through the respective instrument Data Reduction systems, yielding 1D extracted, wavelength calibrated and rectified spectra. The normalization of the continuum was refined with the IRAF task continuum.

## METHODOLOGY

### Line List

We have assembled our Fe I and Fe II line list from [15], [16], [17], plus a selection of lines from the Vienna Atomic Line Database (VALD, [18]). We have, then, visually inspected each line on the observed spectra, in order to check their profile and to discard blended lines. Our final list includes 275 Fe I lines and 37 Fe II lines, spanning the spectral range covered by our FEROS spectra. Our UVES spectra cover a narrower spectral range for which we can use 217 Fe I lines and 30 Fe II lines.

### Equivalent Widths

Several independent and manual (i.e. with the IRAF splot task) measurements of the EW of the whole set of Fe I and Fe II lines in a selected number of Cepheids were performed in order to test the reproducibility of our measures. Our final EW measurements were derived by using a semi-interactive routine developed by one of us (PF, *fitline*). For the determination of the metallicities, we have selected only lines with equivalent widths between 5 and 150 mÅ.

### Stellar Parameters

We have determined the stellar effective temperature by means of the line depth ratios method described in [19]. The total number of line depth ratios adopted to estimate the temperature ranges from 26 to 32 and from 20 to 28 for Galactic and Magellanic Cepheids, respectively. Microturbulent velocity and gravity were constrained by minimizing the log([Fe/H]) vs. EW slope (using the Fe I abundance) and by imposing the ionization balance, respectively.

### Model Atmospheres

We have derived the iron abundances of our stars by using the Kurucz WIDTH9 code [20] and LTE model atmospheres with the new Opacity Distribution Functions computed by [21]. We have used a grid of solar metallicity models for the Galactic and LMC Cepheids and a grid of models computed assuming [Fe/H]=-1.0 and an α-element enhancement of +0.4 dex for the SMC Cepheids [22]. The grids of models have been interpolated in temperature in order to match the effective temperature derived for each star with the line depth ratio method and in steps of 0.10 dex in gravity. In all our computations, we have adopted a solar iron abundance of log[n(Fe)]=7.51 on a scale where log[n(H)]=12 and we have assumed our final stellar metallicity to be the Fe I value, which has been derived by a far larger number of lines with respect to Fe II.

## THE EFFECT OF IRON ON THE PL RELATION

To assess the effect of the iron content on the Cepheid PL relation we select, among our sample, only the stars with periods between 3 and 70 days (61 stars out of 68). We adopted for the barycenter of the LMC a true distance modulus of 18.50±0.1 mag and the SMC is considered 0.44±0.10 mag more distant. For the Galactic stars, we have used two sets of Baade-Wesselink distances ("Old" and "New" in the bottom panels of Figs. 1 and 2). We divide our Cepheid sample into bins of metallicity to investigate its effect on the PL relation. Our results for the V and K bands are presented in Figs. 1 and 2, respectively. In the top panels are shown the PL relations in each metallicity bin calculated with a linear regression. In each panel are also indicated the average iron content of the bin, the root mean square of the linear regression and the number of stars. The bottom panels of Figs. 1 and 2 show the magnitude residuals δ in the V and K band, respectively, as a function of [Fe/H]. In practice, δ(M) is the correction to be applied to a universal, [Fe/H]-independent PL relation to take into account the effects of metallicity. A positive δ means that the actual luminosity of a Cepheid is fainter than the one obtained with the standard PL relation.

Data plotted in the bottom right panel of Fig. 1 indicate that the magnitude residuals of the V band in the individual bins are positive and located at ~2σ (metal-poor) and ~9σ (metal-rich) from zero. Moreover, the δ($M_V$) in the two bins differ from each other at the 3σ level. Our data, then, suggest that metal-rich Cepheids in the V band are, at a fixed period, fainter than metal-poor ones. Adopting different distance determinations to the Galactic stars does not affect this conclusion.

The results for the K band are displayed in Fig. 2. At variance with the V band, in K adopting different distances to the Galactic Cepheids leads to different conclusions: no dependence of the PL on metallicity for the "Old" distances, a 4σ effect for the "New" ones. This latter conclusion would be at odds with current empirical (e.g. [14]) and theoretical (e.g. [23]) evidence suggesting that the PLK relation is, if at all, only marginally affected by metal abundance.

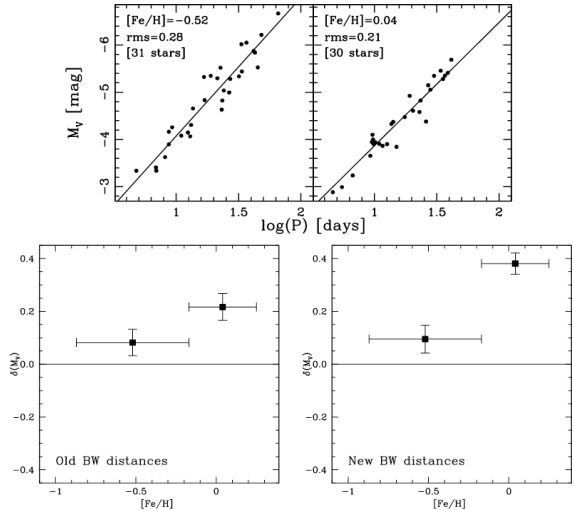

**FIGURE 1.** The top panels show the PL relations calculated in each bin for the V (left figure) and K band (right figure). The bottom panels show the residuals as a function of the iron abundance for both the "Old" and the "New" Galactic Cepheid distances. The mean values in each metallicity bin are plotted as filled squares, with the vertical error-bars representing the associated errors. The horizontal bars indicate the dimension of the bins. The horizontal solid lines indicate the null value that corresponds to an independence of the PL relation from the iron content.

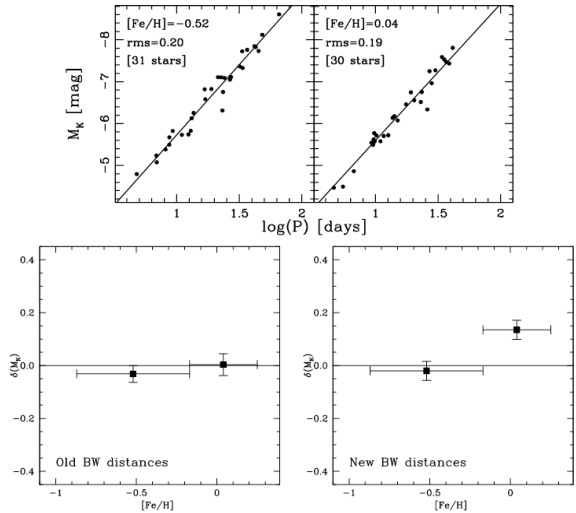

**FIGURE 2.** Same as Figure 1, but in the K band.

## CONCLUSIONS

Our data show a clear increasing trend of the PL residuals in the V band with the iron content (see Fig. 1). This result is in disagreement with an independence of the PL relation on iron abundance and with the linearly decreasing trend found by other observational studies in the literature (e.g. Kennicutt et al. 1998). On the other hand, no firm conclusion can be reached in the K band because the uncertainties in the Galactic distances are large compared to the intrinsic effect of iron on the PL (cf. Figure 2).